\documentclass{iau} 
\usepackage{graphicx}
\usepackage{subcaption}

\title[Exploring the SFH of Galaxies in Different Environments from MaNGA Spectra] 
{Exploring the Star Formation Histories of Galaxies in Different Environments from MaNGA Spectra}

\author[Maria Argudo-Fern\'andez et al.]   
{Maria Argudo-Fern\'andez,$^{1,2}$
 \and M\'ed\'eric Boquien,$^2$
 \and Shiyin Shen,$^3$     
 \and Fangting Yuan,$^3$
 \and Jun Yin,$^3$
 \and Ruixiang Chang,$^3$
 \and Lei Hao$^3$}              

\affiliation{$^1$Centro de Astronom\'ia (CITEVA), Universidad de Antofagasta, Avenida Angamos 601 Antofagasta, Chile\\ 
 email: {\tt maria.argudo@uantof.cl}\\[\affilskip]
$^2$Chinese Academy of Sciences South America Center for Astronomy, China-Chile Joint Center for Astronomy, Camino El Observatorio 1515, Las Condes, Santiago, Chile\\[\affilskip]
$^3$Shanghai Astronomical Observatory, Chinese Academy of Sciences, 80 Nandan Road, Shanghai 200030, China}

\pubyear{2018}
\volume{341}  
\jname{PanModel2018 : Challenges in Panchromatic Galaxy Modelling with Next Generation Facilities}
\editors{M. Boquien, E. Lusso, C. Gruppioni, \& P. Tissera, eds.}
\begin{document}

\maketitle

\begin{abstract}
The star formation history (SFH) of galaxies allow us to investigate when galaxies formed their stars and assembled their mass. We can constrain the SFH with high level of precision from galaxies with resolved stellar populations, since we are able to discriminate between stars of different ages from the spectrum they emit. However, the relative importance of secular evolution (nature) over nurture is not yet clear, and separating the effects of interaction-driven evolution in the observed galaxy properties is not trivial. The aim of this study is to use MaNGA (Mapping Nearby Galaxies at APO) Integral Field Unit (IFU) data, in combination with multi-wavelength data, to constrain the SFH of nearby isolated galaxies. We present here the new techniques we are developing to constrain the SFH with high level of precision from Spectral Energy Distribution (SED) fitting. This study is part of a China-Chile collaboration program where we are applying these new techniques to investigate how galaxies formed and evolve in different environments. 
\end{abstract}

\firstsection 
\section{Introduction}

The electromagnetic emission of galaxies is our main window into their formation and their evolution. Young stellar populations dominate the UV-optical energy budget of galaxies. As they evolve and die they produce dust and inject vast amounts of metals in the interstellar medium, which have a direct impact on the formation of new stars, shielding and cooling molecular clouds, but also reddening and dimming the UV-to-near-infrared spectrum. At the same time, galaxy can be red and dim simply because they are not forming stars anymore. In other words, the electromagnetic spectrum of galaxies encodes the past and current physical processes that have driven their evolution across cosmic times.

One of the major spectroscopic surveys in existence is the Sloan Digital Sky Survey (SDSS; \cite[York et al. 2000]{2000AJ....120.1579Y}, \cite[Gunn et al. 2006]{2006AJ....131.2332G}, \cite[Aguado et al. 2019]{2019ApJS..240...23A}). It has provided us with spectra for a vast number of low redshift galaxies, yielding for each of them information about their star formation history, dust extinction, metallicity, stellar and ionized gas kinematics, etc. However, SDSS spectra are limited to a single spatial position at the center of galaxies (\cite[Strauss et al. 2002]{2002AJ....124.1810S}). The advent of the MaNGA (Mapping Nearby Galaxies at APO, \cite[Bundy et al. 2015]{Bundy+15}) Integral Field Spectroscopy (IFS) survey is a giant leap forward. The survey is one of three core programs in the fourth-generation SDSS (SDSS-IV, \cite[Blanton et al. 2017]{Blanton+17}). It provides us with a 3-dimensional view of galaxies (one spectral, two spatial) that encodes the processes driving their evolution, allowing us to go far beyond either the broadly used approaches of spatially-resolved photometry and spatially-unresolved spectroscopy. 

What we present here is part of a China-Chile collaboration research program. Our aim is to gain a unique insight into how nearby galaxies form and evolve by comparing their observed properties with results from state-of-the-art chemical evolution models. MaNGA 3-D maps allow us to  measure their physical properties (SFH, mass, metallicity, ionized gas and stellar continuum extinctions, etc.) with unprecedented precision and accuracy combining with multi-wavelength spectral energy distribution (SED) modeling from the ultraviolet to the infrared. This will therefore provides us new constraints on the assembly of galaxies across cosmic times. Because processes governing the evolution of galaxies can be strongly intertwined, we collaborate to address these issues putting our own tools and expertise in common, from fundamental spectral modeling to the interpretation of the results of these models to understand the effect of the environment on the evolution of galaxies. In particular, we show here the developed techniques and first results on the study of the evolution of spiral isolated galaxies.

\section{Modelling of MaNGA galaxies}
\subsection{Photometric and spectral modelling}

Spectral and photometric models are outstanding tools for measuring the physical properties of galaxies. Yet, they both have their own strengths and weaknesses. Spectra allow us to detect the presence of an AGN and access to the physical conditions of the ionized gas and of the stellar populations through emission and absorption lines and spectral indices. This yields, for instance, estimates of the extinction of the ionized gas, of the metallicity (both of the ionized gas and the stars), and constraints on the star-formation rate (SFR) on very short (Halpha, $\sim$10~Myr) and very long (D4000, $\sim$1~Gyr) timescales. However spectra are limited to a relatively narrow range in wavelengths and thus are only sensitive to a fraction of the baryonic components of galaxies. Conversely, photometric observations from the UV to the far-IR allow us to also probe young stellar populations that dominate the UV and the dust that reprocesses short wavelength radiation into the IR. Combined together, such a broad wavelength coverage provides us with the leverage we need to constrain the attenuation curve of the stellar continuum, the UV-based and IR-based SFR, and the stellar mass of galaxies. In other words, spectral and photometric data provide us with complementary and synergistic constraints on the physical properties of galaxies.

\subsection{CIGALE}

To perform the SED modelling combining all these data we use the state-of-the-art Code Investigating GALaxy Emission (CIGALE\footnote{\texttt{https://cigale.lam.fr}}, \cite[Boquien et al. 2018]{CIGALE}). CIGALE is a self-consistent SED modelling code from the far-ultraviolet to the radio domain, including the contributions from stellar populations of all ages, thermal and non-thermal gas emission, dust (both in absorption and emission: the energy absorbed by the dust in the UV-to-near-IR domain is re-emitted self-consistently in the mid- and far-IR), and optionally an active nucleus. It allows us to estimate the physical properties of the modelled targets (most importantly SFR over different timescales, burst strength, metallicities, dust attenuation and stellar masses). It is based on a Bayesian-like analysis method. Therefore, used in combination with spatially resolved multi-wavelength data to complement MaNGA observations, this will allow us to obtain better estimates of the SFR, the SFH, and other physical properties of the galaxies in our sample. In this study we use CIGALE to create a set of modelled spectra and use the same Bayesian-like analysis method to fit the observed MaNGA spectra and to extract the physical parameters of the SFH of the stellar populations. 

\subsection{Application to a galaxy merger}

The UV and IR data are usually used to constrain the star formation rate and dust attenuation for mergers, which are tracers of the evolutionary stage of the systems. However, spatially resolved studies of galaxy mergers are restricted to very close galaxies, especially due to the lack spatial resolution in IR data. In \cite{Yuan+18}, we show how the use of UV-to-IR broadband SED, in combination with MaNGA IFS, provide unique opportunity to study the star formation histories and dust attenuation at the tail and core parts of the merging galaxy Mrk 848 (see Figures 2 and 3 in \cite[Yuan et al. 2018]{Yuan+18}). We find that dust derived from full spectra fitting is systematically lower than from UV to IR SED fitting (see Fig.~9 in \cite[Yuan et al. 2018]{Yuan+18}). Additionally, using both photometry and spectroscopy SED modelling, we find a starburst younger than 100~Myr in core region, consistent with the scenario where the interaction-induced gas in-flow enhances the star formation in the center of galaxies.

\begin{figure}
\centering
\includegraphics[width=0.5\textwidth]{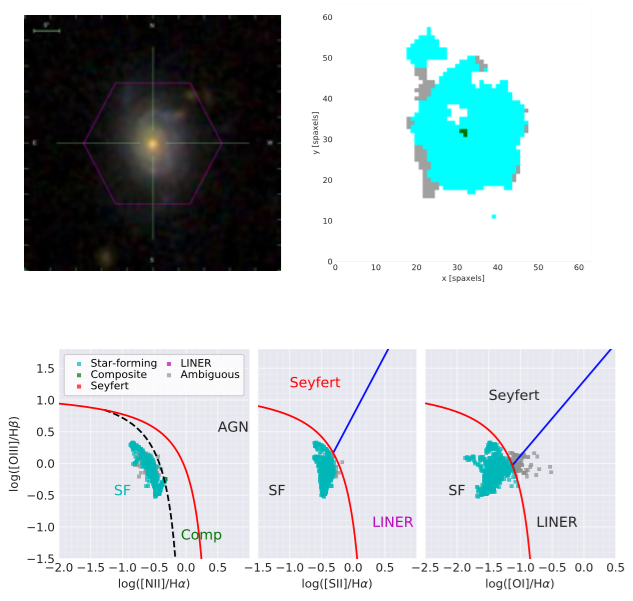}
\includegraphics[trim={0 0 0 1.3cm},clip, width=.45\textwidth]{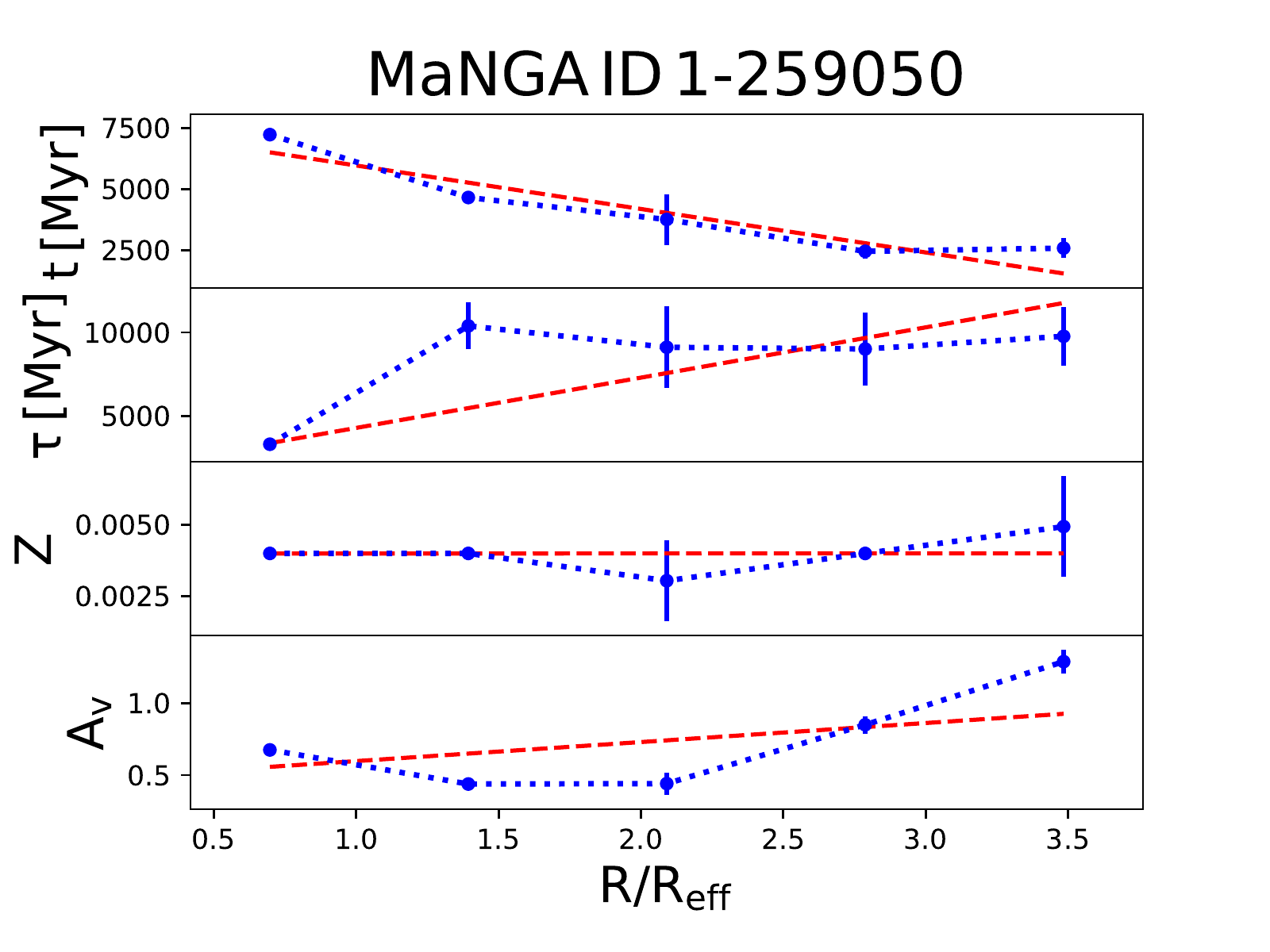}
\caption{Results of the spectroscopic analysis for a SF SIG galaxy (Argudo-Fern\'andez et al., in prep.). \textit{Left panel:} Spatially resolved BPT diagrams and map of the SIG galaxy of the upper left panel, colored as a function of the BPT classification according to the legend. \textit{Right panel:} Bayesian-like estimation of SFH's parameters (age, $\tau$), stellar metallicity (Z), and dust attenuation (A$_{\rm v}$) in radial bins as a function of the effective radius $\rm R_{eff}$.}
\label{Fig:isol}
\end{figure}

\firstsection 
\section{Evolution of spiral isolated galaxies}

We are interested into explore how galaxies build their disk when they grow in isolation, unaffected by external influences. From models, it is predicted that disks grow from inner to outer parts of galaxies, the so-called inside-out model of galaxy formation. As a consequence, we expect to observe gradients in stellar age and metallicity as a function of galaxy radii. The sample of isolated galaxies used in this study is selected from the SDSS-based catalog of Isolated Galaxies (SIG; \cite[Argudo-Fern{\'a}ndez et al.2015]{ArgudoFernandez+15}), composed of 3702 galaxies. Of the $\sim$200 SIG galaxies observed by MaNGA, we selected a sample of 18 almost face-on star-forming (SF) spiral galaxies. We focused on SF galaxies to assure that our results are not contaminating by quenching mechanisms (e.g., LINER or AGN emission). To identify the SF galaxies, we used Marvin (\cite[Cherinka et al. 2018]{Marvin}) to create spatially-resolved BPT diagrams and selected the galaxies where more than the 90\% of their spaxels are classified as star-forming (see the left panel of Fig.~\ref{Fig:isol}). 

Considering that SIG galaxies in our sample are not subject to significant influence from close neighbours for significant time periods, we assume that their disk has homogeneously evolved form inside-out and we therefore divide each SIG galaxy in radial bins with separations of 3\,arcsec (MaNGA spatial resolution is 2.5\,arcsec). We finally use CIGALE, as explained in the previous section, to get the parameters of the SFH (see the right panel of Fig.~\ref{Fig:isol}). We find that, in general, the age $t$ and $\tau$ gradients of spiral SIG galaxies are in agreement in the inside-out disk growth. However, the stellar metallicity (Z) and dust attenuation (A$_{\rm v}$) gradients show more diverse results, which indicate that the evolution of the disk is more complex than expected (Argudo-Fern\'andez et al. in prep.).

\acknowledgments
{\scriptsize MAF acknowledges IAU Travel Grant for S341 and CAS-CONICYT N° 17002 for funding support to assist this conference. Funding for the Sloan Digital Sky Survey IV has been provided by the Alfred P. Sloan Foundation, the U.S. Department of Energy Office of Science, and the Participating Institutions. SDSS-IV acknowledgessupport and resources from the Center for High-Performance Computing at the University of Utah. The SDSS web site is www.sdss.org.}

\newpage
\begin{discussion}

Maria Argudo Fern\'andez\\
Sadanori Okamura\\
Q. Could you please summarise the implication (meaning) of the plots where various parameters are shown with color of the circles representing the radius from the center of a galaxy.\\
A. In the figure we show how the physical parameters of the SFH, derived from the CIGALE spectral fitting, are related each other. This is to check that there is not any degeneracy between the parameters (we create a large grid of modelled spectra to avoid this). We show how the mean age of the stellar population ($t$), the time-scale ($\tau$) parameters of the SFH, the stellar metallicity (Z), and the dust attenuation (A$_{\rm v}$) are related each other. Each point (coloured as a function of the effective radius) corresponds to one binned spectrum for all the considered galaxies. We observe a feature in the relation between the parameters $t$ and $\tau$, which is a direct consequence of the SFH, since we modelled spectra with a physically meaning SFH. \\
==\\
Maria Argudo Fern\'andez\\
Buat Veronique\\
Q. You find an higher attenuation by dust due to SED fitting than from spectral fits, it is at odds with a higher attenuation in emission lines, are you dominated by the spectral continuum?\\
A. Yes, with the spectral fitting we focus on the continuum and the absorption lines. We mask the emission lines and create the grid of models without doing any assumption on the nebular emission.\\
==\\ 
Maria Argudo Fern\'andez\\
Tomotsugu Goto\\
Q. Optical only SED fitting underestimate A$_{\rm v}$. But MANGA is optical only?\\
A. That is why our aim is to extend CIGALE to do photo + spectral fitting all together, to have more info on the UV and far-IR. But optical spectra is important because we can get information from many points in the SED, not only from a few points as usually occurs using photometry only, as we can get many information from the continuum. 

\end{discussion}

\end{document}